\documentclass{iopart}

\usepackage{iopams}
\usepackage{cite}
\usepackage{graphicx}

\newcommand{\bd}{\begin{displaymath}}
\newcommand{\ed}{\end{displaymath}}
\newcommand{\be}{\begin{equation}}
\newcommand{\ee}{\end{equation}}
\newcommand{\ba}{\begin{eqnarray}}
\newcommand{\ea}{\end{eqnarray}}

\begin{document}

\title[Electromagnetic energy flow lines as possible paths of photons]
{Electromagnetic energy flow lines as possible paths of photons}

\author{M Davidovi\'c$^1$, A S Sanz$^2$, D Arsenovi\'c$^3$,
M Bo\v zi\'c$^3$\\ and S Miret-Art\'es$^2$}

\address{$^1$Faculty of Civil Engineering, University of Belgrade,\\
Bulevar Kralja Aleksandra 73, 11000 Belgrade, Serbia \\
$^2$Instituto de F\'{\i}sica Fundamental,\\
Consejo Superior de Investigaciones Cient\'{\i}ficas,\\
Serrano 123, 28006 Madrid, Spain\\
$^3$Institute of Physics, University of Belgrade,\\
Pregrevica 118, 11080 Belgrade, Serbia}

\eads{\mailto{milena@grf.bg.ac.yu},\mailto{asanz@imaff.cfmac.csic.es},
\mailto{arsenovic@phy.bg.ac.yu},\mailto{bozic@phy.bg.ac.yu},
\mailto{s.miret@imaff.cfmac.csic.es}}

\begin{abstract}
Motivated by recent experiments where interference patterns behind a
grating are obtained by accumulating single photon events, here we
provide an electromagnetic energy flow-line description to explain the
emergence of such patterns.
We find and discuss an analogy between the equation describing these
energy flow lines and the equation of Bohmian trajectories used to
describe the motion of massive particles.
\end{abstract}

\pacs{03.50.De, 03.65.Ta, 42.25.Hz, 42.50.-p}





\section{Introduction}
\label{sec1}

The possibility of performing quantum interference experiments with
low-intensity beams (i.e., one per one particle) of photons
\cite{Parker,Dimitrova,Weiss} and material particles
\cite{Rauch,Tonomura,Shimizu} has intensified the theoretical search
of the topology of the photon paths \cite{Prosser,Ghose,Holland} and
particle trajectories \cite{Sanz-1,Gondran,Sanz-2,Bozic-1,Bozic-2}
that describe the process behind the interference grating.
The aim of all the proposed approaches is to simulate the appearance of
the interference pattern by accumulation of single-particle events.

In Bohmian mechanics one may simulate this process for material
particles.
Bohmian trajectories follow the streamlines associated with the
quantum-mechanical probability current density and, therefore,
reproduce exactly the quantum-mechanical particle space distribution
in both the near and the far fields \cite{Sanz-1,Gondran,Sanz-2}.
Alternatively, the emergence of the interference pattern in the far
field has also been simulated by sets of rectilinear trajectories
characterized by the momentum distribution associated with the particle
wave function \cite{Bozic-1,Bozic-2}.
In the far field, the distribution of momentum components along Bohmian
trajectories is consistent with this distribution \cite{Bozic-2}.

In this paper, we show how to determine electromagnetic energy
(EME) flow lines behind an interference grating, where the components
of the magnetic and electric vector fields satisfy Maxwell's equations.
These fields are expressed in terms of a function which explicitly
takes into account the boundary conditions imposed by the grating.
The EME flow lines are then determined after numerically solving the
path equation arising from the Poynting energy flow vector.
In particular, here we show EME flow lines behind gratings consisting
of different number of slits.
These sets of lines supplement those presented by Prosser
\cite{Prosser} for both a semi-infinite plane and gratings with one
and two slits.
The EME flow lines show that the energy redistribution behind the
grating until reaching the Fraunhofer regime.
In particular, it is of interest the process that corresponds to
multiple-slit gratings, where we can observe the smooth transition from
a Talbot pattern in the near field to the characteristic Fraunhofer
peaks in the far field.

It is tempting to conclude from the results obtained that the motion of
an eventual photon wave packet thus represents an energy flow along a
group of flow lines.
This conclusion is supported also by the fact that the path equation
for the EME flow lines has the same form as the equation for the
quantum flow associated with material particles.
This explains why there is a complete similarity in interference
phenomena with photons and material particles.
Experimentally, the final interference patterns as well as the
processes of their emergence are analogous \cite{Parker,Dimitrova,%
Weiss,Rauch,Tonomura,Shimizu}.


\section{The complex Poynting vector and the equation for the EME flow
lines}
\label{sec2}

The diffraction of electromagnetic radiation by a grating is described
by the solution of Maxwell's equations in vacuum that satisfy the
grating boundary conditions \cite{Sommerfeld}.
We consider the simplest solutions of Maxwell's equations: harmonic
electromagnetic waves
\ba
 \tilde{\bf H}({\bf r},t) & = & {\bf H}({\bf r}) e^{-i\omega t} ,
 \label{eq1} \\
 \tilde{\bf E}({\bf r},t) & = & {\bf E}({\bf r}) e^{-i\omega t} .
 \label{eq2}
\ea
The physical electric and magnetic fields are obtained by taking the
real parts of the corresponding complex quantities.
The space-dependent parts of these fields (which are complex
amplitudes) satisfy the time-independent Maxwell equations
\ba
 \nabla \times {\bf H}({\bf r}) & = &
  - i\omega\epsilon_0 {\bf E}({\bf r}) ,
 \label{eq3} \\
 \nabla \times {\bf E}({\bf r}) & = &
  i\omega\mu_0 {\bf H}({\bf r}) ,
 \label{eq4} \\
 \nabla \cdot {\bf H}({\bf r}) & = & 0 ,
 \label{eq5} \\
 \nabla \cdot {\bf E}({\bf r}) & = & 0 .
 \label{eq6}
\ea
From these equations it follows that both fields, ${\bf E}({\bf r})$
and ${\bf H}({\bf r})$, satisfy the Helmholtz equation
\ba
 \nabla^2 {\bf H}({\bf r}) + k^2 {\bf H}({\bf r}) & = & 0 ,
 \label{eq7} \\
 \nabla^2 {\bf E}({\bf r}) + k^2 {\bf E}({\bf r}) & = & 0 ,
 \label{eq8}
\ea
where $k = \omega/c$.

The EME flow lines are now determined from the energy flux vector,
which is the time-averaged flux of energy, given by the real part of
the complex Poynting vector \cite{Jackson}
\be
 {\bf S}({\bf r}) = \frac{1}{2} \ \! {\rm Re}[{\bf E}({\bf r})
  \times {\bf H}^*({\bf r})] .
 \label{eq9}
\ee
Note that, since the flow of energy goes in the direction of the
Poynting vector, the EME flow lines can then be determined from the
parametric differential equation
\be
 \frac{d{\bf r}}{ds} = \frac{1}{c}\frac{{\bf S}({\bf r})}{U({\bf r})} ,
 \label{eq10}
\ee
were $s$ is a certain arc-length along the corresponding path and
$U({\bf r})$ is the time-averaged electromagnetic energy density
\be
 U({\bf r}) = \frac{1}{4} \ \!
  [\epsilon_0 {\bf E}({\bf r}) \cdot {\bf E}^*({\bf r})
   + \mu_0 {\bf H}({\bf r}) \cdot {\bf H}^*({\bf r})] .
 \label{eq11}
\ee


\section{The Poynting vector in the case of two dimensional diffraction
by a plane grating}
 \label{sec3}

To describe simple diffraction experiments, we will consider that the
grating is on the $XZ$ plane at $y = 0$, with the incident plane
harmonic wave traveling along the $y$-direction.
Moreover, we assume the problem to be completely independent of the
$z$-coordinate (i.e., very long slits along this coordinate).
In order to encompass all possible cases of polarization, we express
the magnetic and electric fields before the grating as a superposition
of two waves: $H$ polarized, for which the magnetic field is along $z$
($A$ components), and $E$ polarized, for which the electric field is
along $z$ ($B$ components) \cite{Born}.
That is,
\ba
 \tilde{\bf H}({\bf r},t) & = &
  {\bf H}({\bf r}) e^{-i\omega t} =
  \left[ A e^{iky} {\bf e}_z + B e^{i\phi} e^{iky} {\bf e}_x\right]
   e^{-i\omega t} ,
 \label{eq12} \\
 \tilde{\bf E}({\bf r},t) & = &
  {\bf E}({\bf r}) e^{-i\omega t} = \frac{k}{\epsilon_0\omega}
  \left[ - A e^{iky} {\bf e}_x
    + B e^{i\phi} e^{iky} {\bf e}_z\right] e^{-i\omega t} .
 \label{eq13}
\ea
Here, $\phi$ is the phase shift between the $x$-and $z$-component of
the field, and the constants $A$ and $B$ are real.
For $\phi=0$ or $\pi$, the incident wave is linearly polarized, while
the cases $A=B$, with $\phi=\pm \pi/2$, describe circular polarization.
The cases $A \ne B$, with $\phi \ne 0, \pi$ and $\pm \pi/2$, describe
elliptic polarization.

As shown by Born and Wolf \cite{Born}, the $E$-polarization and
$H$-polarization components satisfy two independent sets of equations
since the problem is independent of the $z$-coordinate.
This implies that the solution that corresponds to an incident wave,
given by (\ref{eq12}) and (\ref{eq13}), diffracted by a grating can be
expressed as
\ba
 {\bf H} & = & i k^{-1} B e^{i\phi} \ \!
  \frac{\partial \psi}{\partial y} \ \! {\bf e}_x
 - i k^{-1} B e^{i\phi} \ \!
  \frac{\partial \psi}{\partial x} \ \! {\bf e}_y
 + A \psi \ \! {\bf e}_z ,
 \label{eq14} \\
 {\bf E} & = & \frac{iA}{\epsilon_0\omega}
  \frac{\partial \psi}{\partial y} \ \! {\bf e}_x
 - \frac{iA}{\epsilon_0\omega}
  \frac{\partial \psi}{\partial x} \ \! {\bf e}_y
 + \frac{k B}{\epsilon_0\omega} \ \! e^{i\phi} \psi \ \! {\bf e}_z ,
 \label{eq15}
\ea
where $\psi(x,y)$ is a solution of the Helmholtz equation
\be
 \nabla^2 \psi + k^2 \psi = 0 ,
 \label{eq16}
\ee
which satisfies the grating boundary conditions.
From (\ref{eq14}) and (\ref{eq15}), we can now express the components
of the time-averaged Poynting vector in terms of the constants $A$, $B$
and $\phi$ and the function $\psi(x,y)$, as
\ba
 S_x & = & \frac{i}{4\epsilon_0\omega} \ \! (A^2 + B^2)
  \left( \psi \frac{\partial \psi^*}{\partial x}
    - \psi^* \frac{\partial \psi}{\partial x} \right) ,
 \label{eq17} \\
 S_y & = & \frac{i}{4\epsilon_0\omega} \ \! (A^2 + B^2)
  \left( \psi \frac{\partial \psi^*}{\partial y}
    - \psi^* \frac{\partial \psi}{\partial y} \right) ,
 \label{eq18} \\
 S_z & = & \frac{i}{2\epsilon_0\omega k} \ \! AB \sin \phi
  \left( \frac{\partial \psi}{\partial x}
    \frac{\partial \psi^*}{\partial y}
    - \frac{\partial \psi^*}{\partial x}
    \frac{\partial \psi}{\partial x} \right) .
 \label{eq19}
\ea

In the case of a linearly polarized initial wave ($\phi = 0$ or $\pi$),
it follows from equation (\ref{eq19}) that the $z$ component of the
Poynting vector vanishes.
Thus, the EME flow lines will remain confined on the $XY$ plane
and, from the parametric differential equation (\ref{eq10}), we find
that the differential equation
\be
 \frac{dy}{dx} = \frac{S_y}{S_x}
  = \frac{\displaystyle \left(\psi\frac{\partial \psi^*}{\partial y}
   - \psi^*\frac{\partial \psi}{\partial y}\right)}
    {\displaystyle \left(\psi\frac{\partial \psi^*}{\partial x}
   - \psi^*\frac{\partial \psi}{\partial x}\right)}
 \label{eq23}
\ee
will determine the eventual photon paths, which are obtained by
numerical integration.
As can be noticed, the topology of the EME flow lines in the case of
linear polarization is independent of the constants $A$ and $B$, and
therefore independent of the direction of polarization.

On the other hand, for circular and elliptic polarizations the
$z$-component of the Poynting vector is non zero, this leading to
important differences in the properties of the corresponding EME
flow lines, which will not be planar, as infers from the non
vanishing $z$-component in equation (\ref{eq10}).
The properties of EME flow lines in these cases are beyond the scope of
the present work and will be explored in a forthcoming one.


\section{Flow lines behind a specific grating}
 \label{sec4}

In order to plot flow lines for a specific grating we need explicit
expressions for the magnetic and electric field behind a grating.
Traditionally, explicit solutions have been written using the solution
of the Helmholtz equation in the form of the Fresnel-Kirchhoff integral
\cite{Born}.
By making the appropriate approximations, the solution was then
transformed into the expressions valid in the Fresnel and Fraunhofer
regions, respectively.
Talbot effect and Talbot-Laue effect were also explained \cite{Clauser}
by making the appropriate approximations and transformations of the
Fresnel-Kirchhoff integral.

If the $x$ component of the wave vector satisfies the relation $k \gg
k_x$, the solution of the Helmholtz equation may also be expressed as
a superposition of transverse modes (STM) of the field multiplied by
an exponential function of the longitudinal coordinate \cite{Bozic-3}.
As shown by Arsenovi\'c \etal \cite{Bozic-4}, this form is equivalent to
the Fresnel-Kirchhoff integral.
For an incident plane wave falling on the grating at $y = 0$ and lying
on the $XZ$ plane, the STM form of the solution reads as
\be
 \psi(x,y) = \frac{e^{iky}}{\sqrt{2\pi}}
  \int_{-\infty}^\infty c(k_x) \ \!
   e^{ik_x x} e^{-ik_x^2 y/2k} dk_x ,
 \label{eq24}
\ee
where the function $c(k_x)$ is determined by the incident wave and the
grating boundary conditions as
\be
 c(k_x) = \frac{1}{\sqrt{2\pi}}
  \int_{-\infty}^\infty \psi(x,0) \ \! e^{-ik_x x} dk_x .
 \label{eq25}
\ee
Within the approximation $k \gg k_x$, one finds
\be
 \left\arrowvert \frac{\partial \psi}{\partial x} \right\arrowvert
 \ll
 \left\arrowvert \frac{\partial \psi}{\partial y} \right\arrowvert ,
 \qquad
 \frac{\partial \psi}{\partial y} \approx ik\psi ,
 \label{eq22}
\ee
and, therefore, the EME density (\ref{eq11}) becomes proportional
to $|\psi(x,y)|^2$, i.e.,
\be
 U({\bf r}) = \frac{1}{2} \ \! \mu_0 (A^2 + B^2) |\psi(x,y)|^2 ,
 \label{eq20}
\ee
which is obtained from (\ref{eq14}), (\ref{eq15}) and (\ref{eq11}).

Assuming that the grating is completely transparent inside the slits
and completely absorbing outside them, we have that $\psi(x,0)=0$
for points outside the slit aperture and $\psi(x,0) = \psi_{in}(x,0)$
for the inside points, where $\psi_{in}(x,0)$ is the field component of
the incident wave.
In the case of an incident plane wave propagating along the $y$-axis,
$\psi_{in}(x,0)$ is a constant.
For a grating with $N$ apertures separated a distance $d$ and all with
the same width $\delta$, a simple integration \cite{Bozic-3} renders
\be
 c(k_x) = \frac{1}{\sqrt{2\pi}} \sqrt{\frac{\delta}{N}}
  \frac{\sin (k_x\delta/2)}{k_x\delta/2}
  \frac{\sin (N k_x d/2)}{\sin (k_x d/2)} .
 \label{eq26}
\ee

\begin{figure}
 \begin{center}
 \includegraphics[width=9cm]{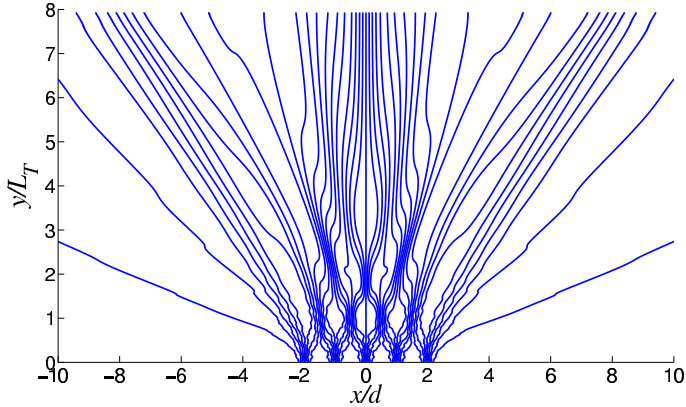}
 \caption{\label{fig1}
  EME flow lines behind a Ronchi grating ($\delta = d/2$) with $N = 5$
  slits, $\lambda = 500$~nm, $d = 20\lambda = 10$~$\mu$m and
  $L_T = d^2/\lambda = 200$~$\mu$m.}
 \end{center}
\end{figure}

\begin{figure}
 \begin{center}
 \includegraphics[width=8cm]{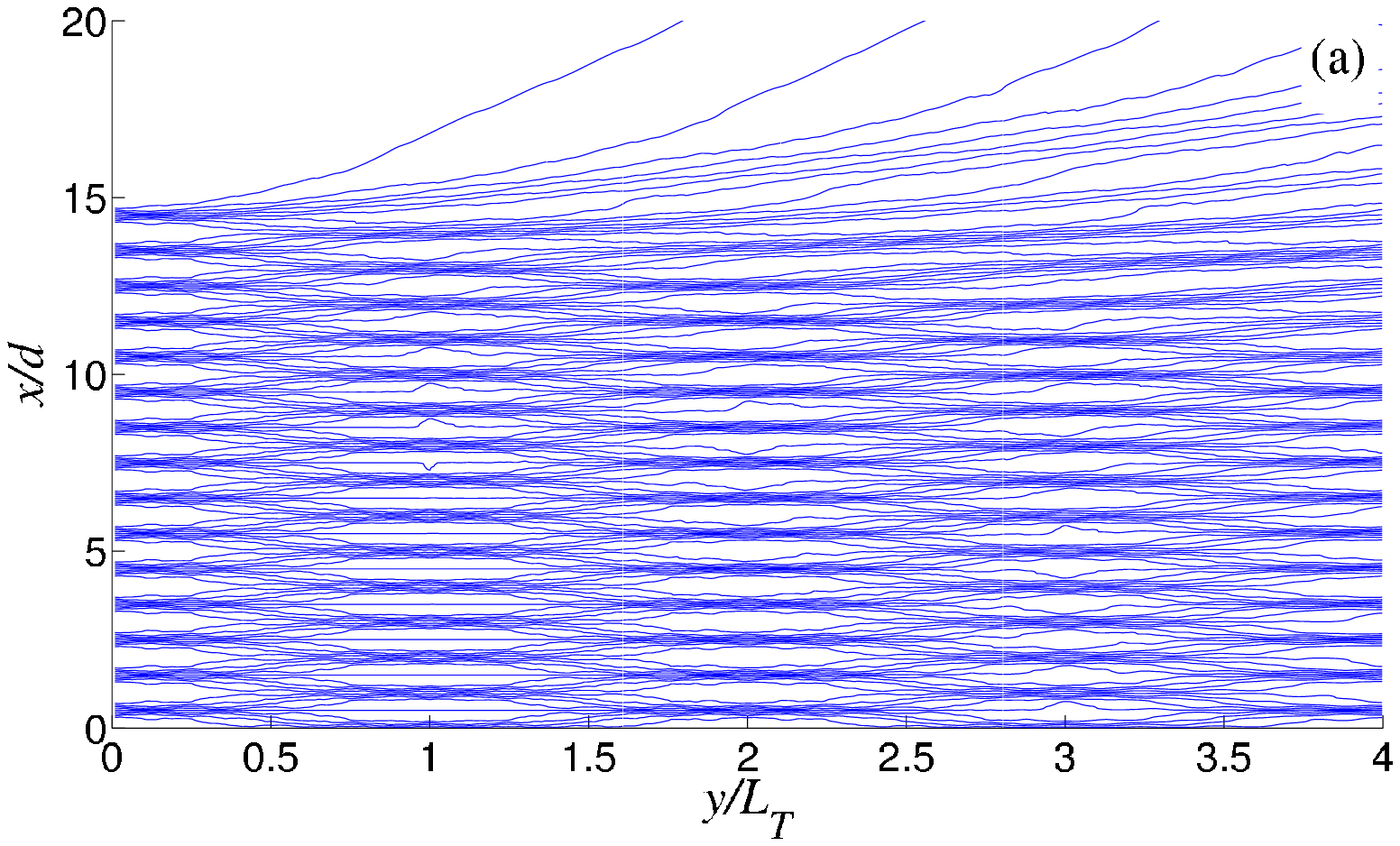}
 \includegraphics[width=8cm]{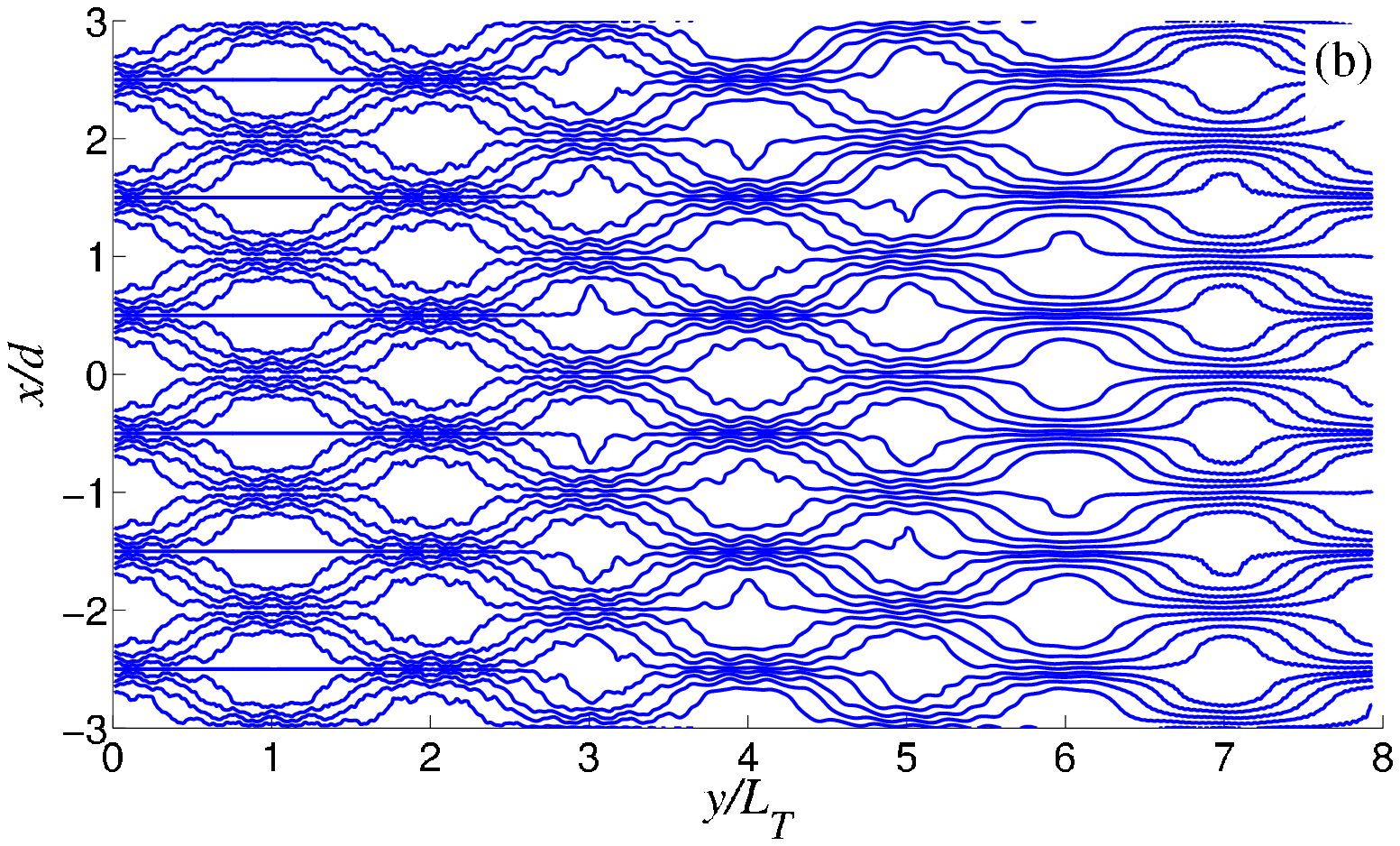}
 \caption{\label{fig2}
  EME flow lines behind a Ronchi grating ($\delta = d/2$) with $N = 30$
  slits, $\lambda = 500$~nm, $d = 20\lambda = 10$~$\mu$m and
  $L_T = d^2/\lambda = 200$~$\mu$m.
  (a) EME lines from the upper part of the grating ($y > 0$).
  (b) EME lines from six central slits of the grating.}
 \end{center}
\end{figure}

In figures \ref{fig1} and \ref{fig2}, we have represented the EME flow
lines behind Ronchi gratings with 5 and 30 slits, respectively.
The unit along the $y$-axis is the so-called Talbot distance,
$\L_T = d^2/\lambda$, which gives the repetition period behind a
grating of the diffracted wave \cite{Sanz-2}; the unit along the axis
parallel to the grating ($x$-axis) is the period $d$ this grating.
As seen, the topology displayed by the EME flow lines is very similar
to that displayed by Bohmian trajectories for massive particles
\cite{Sanz-1,Gondran,Sanz-2}.
Note that the quantum density current
\be
 {\bf J} = \frac{i\hbar}{2m} \ \!
  (\Psi \nabla \Psi^* - \Psi^* \nabla \Psi)
 \label{eq27}
\ee
determines the particle trajectories in Bohmian mechanics through the
equation of motion
\be
 \frac{d{\bf r}}{dt} = \frac{{\bf J}}{\rho} ,
\ee
where $\rho = \Psi^* \Psi$.
The latter equation is analogous to equation (\ref{eq10}), though this
analogy is not totally complete: in the case of particles with a mass
$m$, the relation $y = \hbar k t/m$ holds, whereas the same is not true
for photons.
When the wave function $\Psi$ does not depend on $z$, ${\bf J}$ has
only $x$ and $y$ components,
\ba
 J_x = \frac{i\hbar}{2m} \left(\Psi \frac{\partial \Psi^*}{\partial x}
   - \Psi^* \frac{\partial \Psi}{\partial x}\right)
 \label{eq28} \\
 J_y = \frac{i\hbar}{2m} \left(\Psi \frac{\partial \Psi^*}{\partial y}
   - \Psi^* \frac{\partial \Psi}{\partial y}\right) .
 \label{eq29}
\ea
Equations (\ref{eq28}) and (\ref{eq29}) have the same form as equations
(\ref{eq17}) and (\ref{eq18}).
Thus, the EME flow lines in the case of a linearly polarized incident
wave are similar to the Bohmian trajectories of massive particles
described by a two-dimensional wave function.


\section{Emergence of the interference pattern by accumulation of
single photon arrivals}
 \label{sec5}

In the case of massive particle the modulus square of the wave function
describes the distribution of particles at a distance $y$ from the
grating in the far field after (theoretically) an infinite number of
particles has reached the detector (at $y$).
This theoretical result has been nicely confirmed by a new generation of
experiments which use low-intensity beams of particles.
In these experiments, the final interference patterns are built-up
after particles accumulate gradually one by one at a scanning screen
\cite{Dimitrova,Weiss,Tonomura,Shimizu}.
Numerical simulation of particle arrivals, assuming that they move
along de Broglie-Bohm's trajectories \cite{Sanz-1,Gondran,Sanz-2} and
MD trajectories \cite{Bozic-1} describe theoretically this process.
This means that a trajectory-based interpretation completes the
standard interpretation of the wave function, where a picture in
terms of single events is missing.

\begin{figure}
 \begin{center}
 \includegraphics[width=6.25cm]{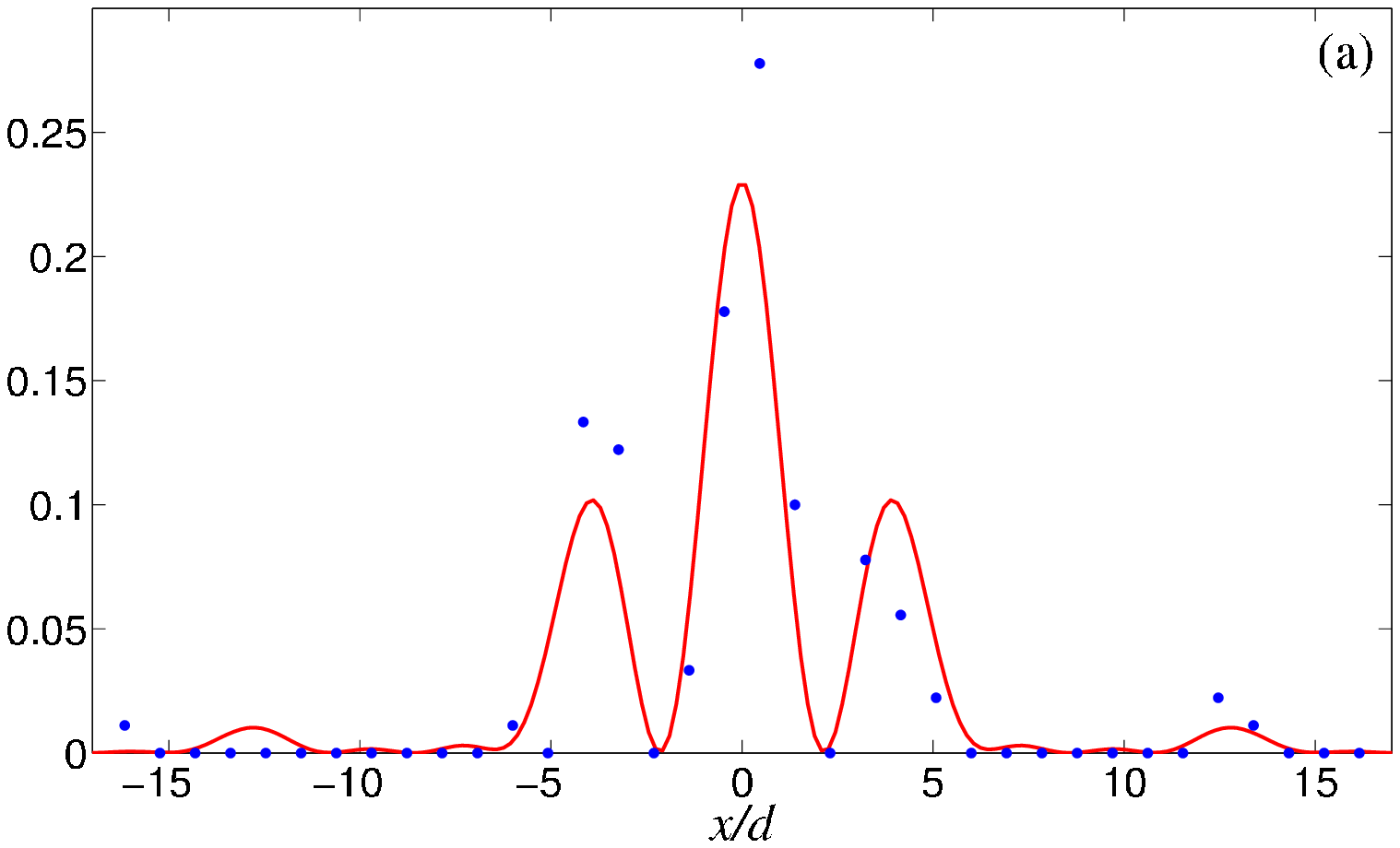}
 \includegraphics[width=6.25cm]{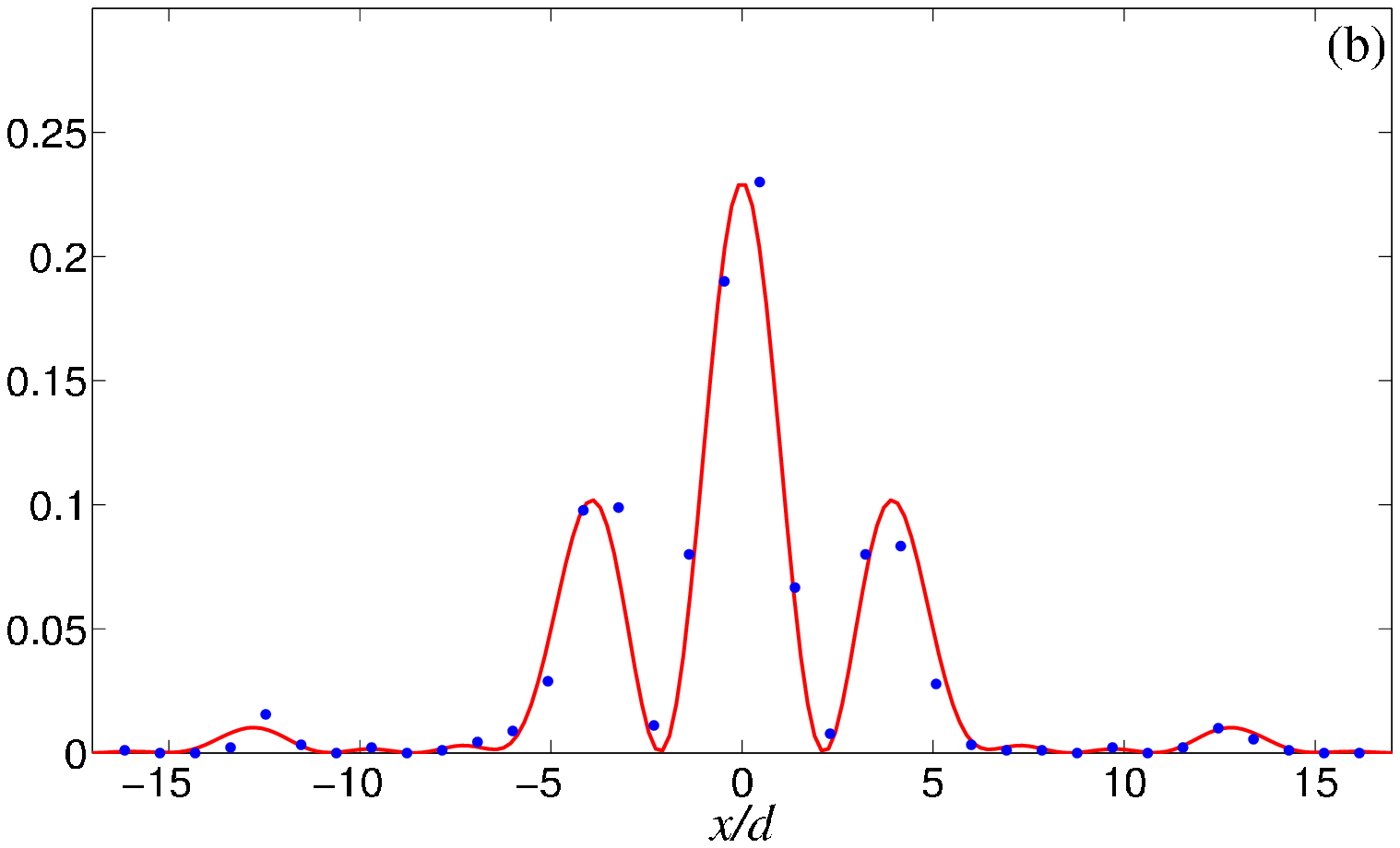}
 \includegraphics[width=6.25cm]{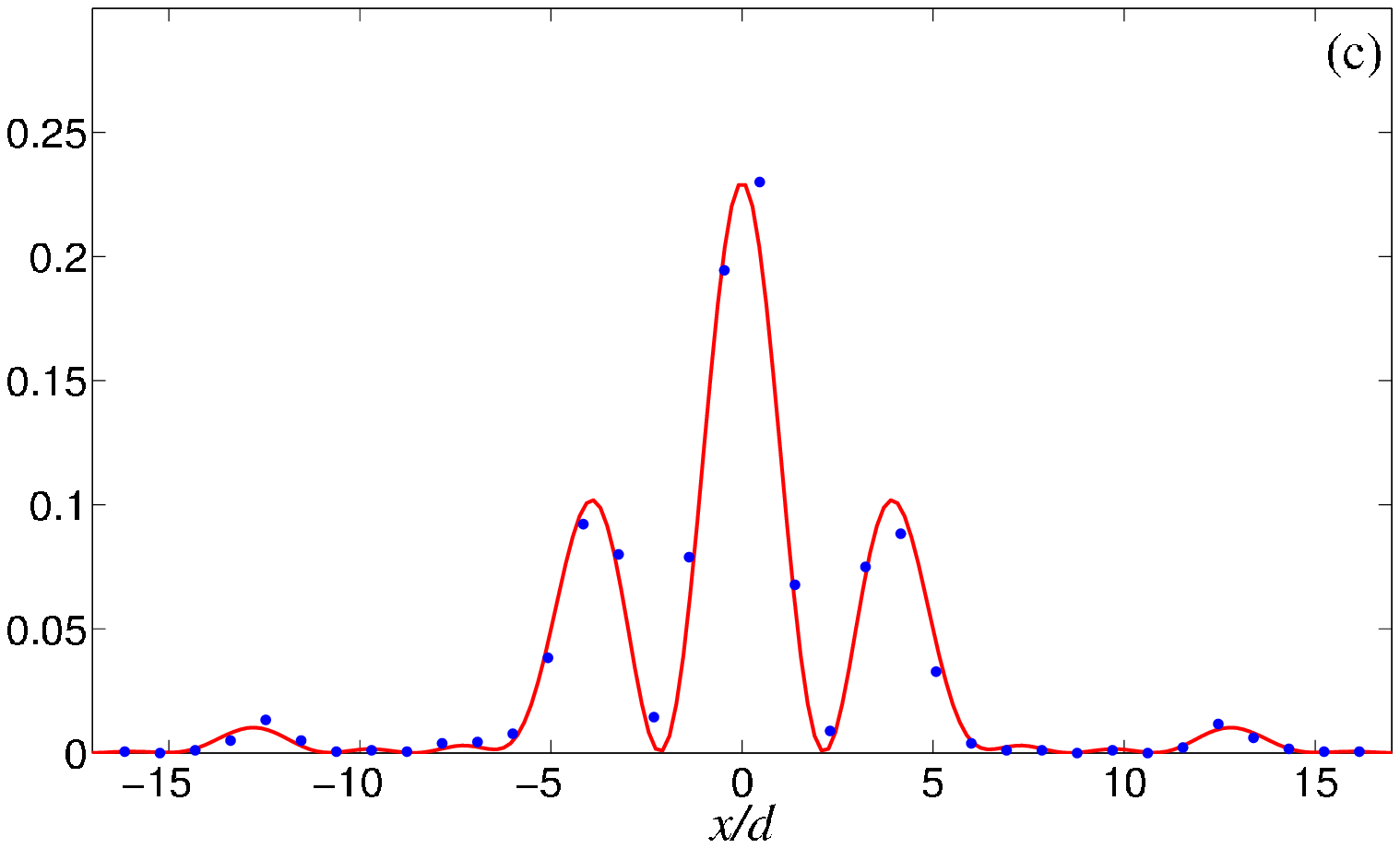}
 \includegraphics[width=6.25cm]{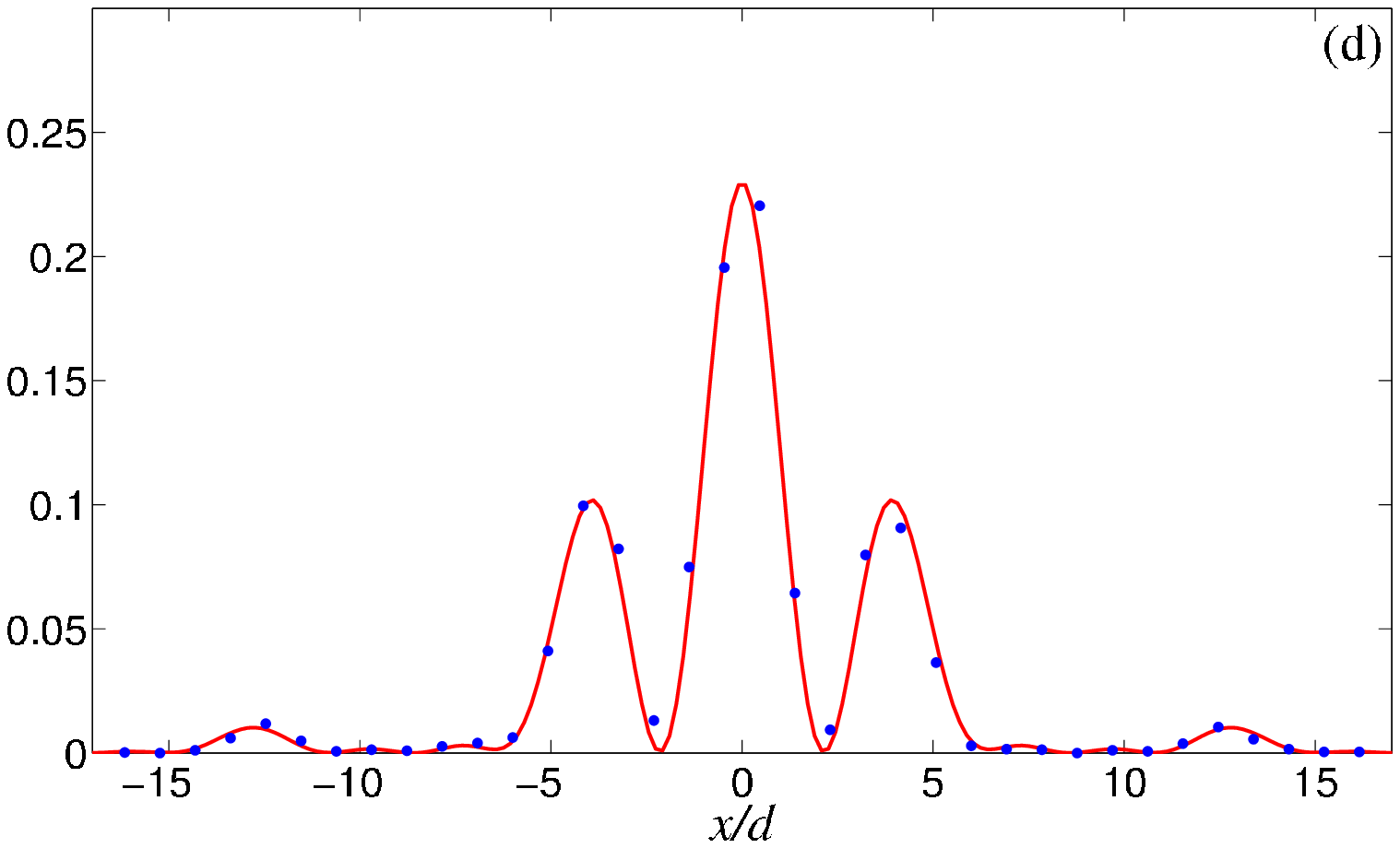}
 \caption{\label{fig3}
  Histogram of the number of trajectories ending at various points
  along the $x$-axis at a distance $y = 4.3L_T$ for four different
  values of the total number of photons: (a) 100, (b) 1000, (c) 2000
  and (d) 5000.
  Here diffraction is produced by a two-slit grating and the initial
  conditions (positions along the two slits) for the photons data are
  chosen at random.
  The red line is the plot of the function $|\psi(x,y)|^2$.
  It is seen that at the chosen distance $y=4.3L_T$, maximum of the
  distribution at $x/d=8.6$, associated with $k_xd=4\pi$, is absent.
  This follows from (25) because for $N=2$, $d=2\delta$, the second
  interference maximum, being at $k_xd=4\pi$ coincides with the first
  diffraction minimum. }
 \end{center}
\end{figure}

Analogously, one can proceed in the same way with photons assuming that
they move along the EME flow lines described in the previous section.
This is illustrated in figure \ref{fig3}, where we have plotted
histograms (blue dots) obtained from the accumulation of photons for
a two-slit diffraction experiment (with $\lambda = 500$~nm, $d = 20
\lambda$ and $\delta = d/2$).
In particular, the histograms have been made considering equally spaced
bins (with a width of $0.9d$) at an observation distance $y_{ob} =
4.3L_T$ from the plane of the grating, where the Fraunhofer pattern is
already well converged (this happens when the observation distance is
further away than the so-called Rayleigh distance \cite{Sanz-3}, which
in this case is $y_R \sim (2\delta + d/2)^2/4\pi\lambda \approx 0.2L_T
\ll y_{ob}$).
As can be seen, as we move from panel (a) to (d) the histogram data
approach better and better the smooth (red) line which represents the
EME density $U(x,y_{ob})$ given by equation (\ref{eq20}) as the number
of photons per time unit increases, as also seen in the experiment
\cite{Dimitrova,Weiss}.
It is interesting to note that the photons (or, equivalently, the
initial positions of their paths) distribute randomly along the
distance covered by each slit aperture and, therefore, their arrival
positions at $y=4.3L_T$ will also be random.
However, they will accumulate in accordance to $U({\bf r})$ because of
the guidance condition given by equation (\ref{eq10}), which can be
alternatively expressed \cite{Born} as
\be
 {\bf S}({\bf r}) = U({\bf r}){\bf v} ,
\ee
where ${\bf v}$ is a sort of effective vectorial velocity field which
{\it transports} the EME density through space in the form of the EME
density current.
The vector field ${\bf v}$ is always oriented in the direction of the
wave vector ${\bf k}$.
Thus, before the grating it is aligned along the $y$ direction and
behind the grating will depend on the particular point $(x,y)$ where
the field is evaluated, becoming almost constant along some specific
direction only within the Fraunhofer regime.


\ack

MD, DA and MB acknowledge support from the Ministry of Science of
Serbia under Project `Quantum and Optical Interferometry', number
141003; ASS and SM-A acknowledge support from the Ministerio de Ciencia
e Innovaci\'on (Spain) under Project FIS2007-62006.
ASS also thanks the Consejo Superior de Investigaciones Cient\'{\i}ficas for
a JAE-Doc Contract.


\end{document}